\documentclass[a4paper, 12pt]{article}
\usepackage{a4}
\usepackage{amsfonts}
\usepackage{amssymb}
\usepackage{epsfig}
\usepackage{psfig}

\newcommand{\be}{\begin{equation}}
\newcommand{\ee}{\end{equation}}
\newcommand{\bea}{\begin{eqnarray}}
\newcommand{\eea}{\end{eqnarray}}

\newcommand{\ti}{\times}
\newcommand{\half}{\frac{1}{2}}

\newcommand{\mc}{\mathcal}

\begin{document}
\linespread{1.2}

\title{
\begin{flushright} \vspace{-2cm}
{\small DAMTP-2005-78 \\ \vspace{-0.35cm}
hep-th/0509012} \end{flushright}
\vspace{2.5cm}
{\bf K\"ahler Moduli  Inflation} 
}

\author{}
\date{}

\maketitle

\begin{center}
Joseph P. Conlon \footnote{e-mail: J.P.Conlon@damtp.cam.ac.uk}
and
Fernando Quevedo \footnote{e-mail: F.Quevedo@damtp.cam.ac.uk} \\
\vspace{1cm}
\emph{DAMTP, Centre for Mathematical Sciences,} \\
\emph{Wilberforce Road, Cambridge, CB3 0WA, UK} \\
\end{center}

\begin{abstract}
\noindent

We show that under general conditions there is at
least one natural inflationary direction
for the K\"ahler moduli of type IIB flux compactifications. This
requires a Calabi-Yau which has $h^{2,1}>h^{1,1}>2$ and for which 
the structure of the scalar
potential is as in the recently found exponentially large volume 
compactifications. We also need - although these conditions may be
relaxed - at least one K\"ahler
modulus whose only non-vanishing triple-intersection is with
itself and which appears by itself in the non-perturbative
superpotential.
Slow-roll inflation then occurs without a fine
tuning of parameters, evading the $\eta$ problem of F-term
inflation. In order to obtain COBE-normalised density perturbations, 
the stabilised volume of the 
Calabi-Yau must be 
$\mc{O}(10^5-10^7)$ in string units, 
and the inflationary scale $M_{infl}\sim 10^{13}$ GeV.
We find a robust  model independent prediction for the spectral
index of $1 - \frac{2}{N_e} = 0.960 \to 0.967$, depending on the number of
efoldings.

\end{abstract}

\thispagestyle{empty}
\clearpage

\section{Introduction}
\label{seINT}

One of the most exciting recent developments in string theory has been
the great progress made in moduli stabilisation \cite{hepth0105097,
  hepth0301240}. There are now well-established techniques to give
masses to the moduli that appear ubiquitously
in string compactifications.
The process of moduli stabilisation represents `step zero' towards string
phenomenology, as the moduli vevs determine such basic quantities as the string scale
and the gauge coupling constants.

Given moduli potentials, an obvious application is to
inflation.
While the moduli sector has only indirect effects on Standard
Model matter, the dynamics of light scalar fields is the principal theme
of inflation. Inflation is the dominant paradigm for structure formation
in the early universe and observations can now provide precision tests
of inflationary models \cite{WMAP}.

In string theory there are several candidates for the inflaton
field, which can be classified according to their origin in either
the open or closed string sector \cite{bg,hepph9812483}. The most
common open string inflaton is a brane/antibrane separation
\cite{hepth0105204, dss, hepth0111025}, whereas closed string
inflatons typically correspond to geometric moduli
\cite{hepth0406230}. There has been much recent effort devoted to
inflationary model building, particularly since the appearance of
the KKLT scenario of moduli stabilisation \cite{hepth0301240}. For
recent discussions, \cite{hepth0406230, hepth0308055, renata,
various} may be consulted.

A standard problem bedevilling both brane and modular inflation -
and indeed most supergravity inflation scenarios - is the $\eta$ problem.
This states that for F-term inflation the
slow-roll $\eta$ parameter is $\mc{O}(1)$ unless a finely tuned
cancellation occurs.
The $\eta$ problem is
manifest for F-term modular inflation. In brane inflation it is
not manifest, but reappears once this is embedded into a moduli stabilisation scenario
\cite{hepth0308055}.

In this note we will present a simple inflationary scenario within
the framework of the moduli stabilisation mechanism of
\cite{hepth0502058, hepth0505076}. The inflaton is one of the K\"ahler
moduli and inflation proceeds by reducing the F-term energy. The
$\eta$ problem is evaded by the pseudo-no scale
property of the K\"ahler potential.
The structure of the potential is such that inflation is obtained
naturally and almost inevitably, without either fine tuning or a need to introduce
large flux or brane numbers. This mechanism in principle applies to a very
large class of Calabi-Yau compactifications that will be specified
below.

\section{Almost Flat Directions}
\label{seGI}

\subsection{General Idea}
\label{sseBPL}

Slow-roll inflation requires the presence of almost flat directions in
the scalar potential.
A natural source of such a flat direction would be a field only appearing
exponentially in the potential. Denoting this field by $\tau$, an
appropriate (and textbook \cite{LiddleLyth}) potential would be
\be
V_{inf} = V_0 \left( 1 - A e^{-\tau} + \ldots \right),
\ee
where the dots represent higher exponents.

In string theory there are many moduli whose stabilisation requires
nonperturbative effects. Examples are the K\"ahler moduli in IIB flux
compactifications and both dilaton and K\"ahler moduli in
heterotic Calabi-Yau compactifications.
We regard all such fields as candidate inflatons,
but shall focus on the K\"ahler
moduli ($T_i$) of type IIB flux compactifications.
These only appear
nonperturbatively in the superpotential, which takes the
form
\be
\label{eqNP}
W = \int G_3 \wedge \Omega + \sum_i A_i e^{-a_i T_i},
\ee
where $T_i = \tau_i + i c_i$ with $\tau_i$ the 4-cycle volume and
$c_i$ the axionic component. The $A_i$ represent threshold corrections
and are independent of the K\"ahler moduli.

Of course, it is well known that $\mc{N} = 1$ F-term inflation suffers from
an $\eta$ problem. Both the K\"ahler potential and
superpotential enter into the scalar potential, and for
generic potentials $\eta \sim \mc{O}(1)$.
However, the key word here is `generic', and the K\"ahler potentials
arising from string theory are (by definition) not generic.
A common way these potentials fail to be generic is by being no-scale,
corresponding to
\be
K^{i \bar{j}} \partial_i K \partial_{\bar{j}} K = 3.
\ee
For a constant superpotential $ W= W_0$, a no-scale scalar potential
vanishes:
\be
V_F = e^K \left( K^{i \bar{j}} D_i W D_{\bar{j}} \bar{W} - 3 \vert W
\vert^2 \right) = 0,
\ee
where $D_i W = \partial_i W + (\partial_i K) W$,
with all directions being exactly flat. In type IIB, the tree-level K\"ahler
potential for the size moduli takes the no-scale form
\be
\mc{K} = - 2 \ln (\mc{V} ),
\ee
where $\mc{V}$ is the internal volume.
Suppose we now add nonperturbative
modular dependence into the superpotential as in (\ref{eqNP}). The scalar potential
becomes
\be
\label{eqPT}
V_F = e^K K^{i \bar{j}} \left[ a_i A_i a_j \bar{A}_j e^{-a_i T_i - a_j \bar{T}_j} -
(  (\partial_i K) W a_j \bar{A}_j e^{-a_j \bar{T}_j} + c.c ) \right].
\ee
$T_i$ only appear nonperturbatively along exponentially flat
directions and
it is natural to ask whether this flatness can drive inflation.

While the potential (\ref{eqPT}) is exponentially flat, it
also appears exponentially small. However, this is only true
so long as all $T_i$ fields are large. In the presence of several
K\"ahler moduli the variation of $V$ along the $T_i$ direction is in general
uncorrelated with the magnitude of $V$ - we note this cannot happen in a
one-modulus model.

There are also extra corrections to the potential, arising
both from the breaking
of no-scale behaviour by K\"ahler corrections and from the uplift terms
needed to fine-tune the cosmological constant.
The latter have several possible sources \cite{hepth0301240, hepth0309187,
  hepth0402135} and scale inversely with the volume
\be V_{uplift} \sim \frac{1}{{\mc V}^{\alpha}}, \ee where
$\frac{4}{3} \le \alpha \le 2$. Notice that the uplift encodes its
modular dependence through the overall volume, rather than depending
explicitly on the moduli.
Thus at
constant volume the $T_n$ direction is extremely flat for large
values of $T_n$.

\subsection{Embedding in IIB Flux Compactifications}
\label{ssePP}

While this is promising, inflation in string theory cannot be
isolated from moduli stabilisation, as the methods used to
stabilise the moduli can generate unacceptably large masses for
the inflaton. We now embed the above in reasonably explicit IIB
flux compactifications\footnote{The lack of explicitness lies
principally in the difficulty of knowing whether and what nonperturbative superpotentials
  will be generated on a particular Calabi-Yau.} and in
particular in the moduli stabilisation mechanism of
\cite{hepth0502058, hepth0505076}. (For other recent work on perturbative
corrections in IIB flux compactifications see \cite{hepth0204254, berghaackkors,
 hepth0507131, per, bw}).

For multi-modulus Calabi-Yaus, evaluating the
scalar potential requires expressing the overall volume in terms of
the 4-cycle volumes, which we shall denote by $\tau_i = \textrm{Re}
(T_i)$.
 For illustration, we shall take a
simplified form for the
Calabi-Yau volume,
\bea\label{vol} \mc{V} & = &
\alpha (\tau_1^{3/2} - \sum_{i=2}^n \lambda_i
\tau_i^{3/2}) \nonumber \\
& = & \frac{\alpha}{2 \sqrt{2}} \left[ (T_1 + \bar{T_1})^{3/2} -
\sum_{i=1}^n \lambda_i (T_i + \bar{T}_i)^{3/2} \right]. \eea 
$\tau_1$ controls the overall
volume and $\tau_2, \ldots, \tau_n$ are blow-ups whose only
non-vanishing triple intersections are with themselves. 
$\alpha$ and  $\lambda_i$ are positive constants depending on the
particular model. The
minus signs are necessary as $\frac{\partial^2 \mc{V}}{\partial
T_i
  \partial T_j}$ must have signature $(1, h^{1,1} -1)$ \cite{CandelasDeLaOssa}.
We stabilise the dilaton and complex structure moduli with 
fluxes and take the K\"ahler moduli superpotential to
be\footnote{ More generally we could take  $W = W_0 +
\sum_{i=2}^n A_i e^{-a_{ij} T_j}$, which would alter
the condition (\ref{eqRAT}) in a model-dependent fashion.
As long as the modified form of (\ref{eqRAT}) can be satisfied, the results for
the inflationary parameters are unaffected. 
In general we expect this to be possible, although we note
that there do exist
models, such as the $\mc{F}_{11}$ model of \cite{hepth0404257}, for
which this cannot be achieved.}
 \be
\label{eqSuP}
W = W_0 + \sum_{i=2}^n A_i
e^{-a_i T_i}, \ee
where $a_i = \frac{2 \pi}{g_s N}$.
The K\"ahler potential is \be \label{eqKPA}
\mc{K} = \mc{K}_{cs} - 2 \ln \left[ \alpha \left(\tau_1^{3/2} -
\sum_{i=2}^n \lambda_i \tau_i^{3/2}\right) + \frac{\xi}{2}
\right],
\ee
where $\xi = -\frac{\chi (M)}{2 (2 \pi)^3}$.
We have included the $\alpha'$ corrections of
\cite{hepth0204254}. The dilaton has been fixed and so we can define
the moduli
using either string or Einstein-frame volumes; we use the former.
If the latter, we must replace $a_i \to a_i g_s$
and $\xi \to \xi g_s^{-3/2}$ in (\ref{eqSuP}) and (\ref{eqKPA}) - the physics is of
course the same.
As we work in the moduli stabilisaton framework of
\cite{hepth0502058,
  hepth0505076} we anticipate that at the minimum we will have $\tau_1 >> \tau_i$
and $\mc{V} >> 1$. The resulting scalar potential is
\be
V = e^K \left[ G^{i \bar{j}} \partial_i W \partial_{\bar{j}} \bar{W}
+ G^{i \bar{j}} \left((\partial_i K) W) \partial_{\bar{j}} \bar{W} +
c.c. \right) \right] + \frac{3 \xi W_0^2}{4 \mc{V}^3}.
\ee
We need $\xi > 0$ and so require $h^{2,1} > h^{1,1}$. For the above K\"ahler potential, we have
\be
G^{i \bar{j}} \sim \frac{8 \mc{V} \sqrt{\tau_i}}{3 \alpha \lambda_i} \delta_{ij} + \mc{O}(\tau_i \tau_j).
\ee
$G^{i \bar{j}}$ is real and, up to terms subleading in volume,
 satisfies $G^{i \bar{j}} \partial_{\bar{j}}
  \mc{K} = 2 \tau_i$.
At large volume only the leading part of $G^{i \bar{j}}$ is
relevant and the scalar potential becomes \be \label{eqSPT} V =
\sum_i \frac{8 (a_i A_i)^2 \sqrt{\tau_i}}{3 \mc{V} \lambda_i
\alpha} e^{-2 a_i \tau_i} - \sum_i 4 \frac{a_i A_i}{\mc{V}^2} W_0
\tau_i e^{-a_i \tau_i} + \frac{3 \xi W_0^2}{4 \mc{V}^3}. \ee The minus
sign in the second term arises from setting the $b_i$ axion to its
minimum. There are terms not included in (\ref{eqSPT}), but these
are subleading. 
Importantly, they only depend on $\tau_i$ through the overall volume. 
This is crucial and ensures that at large
$\tau_i$ the variation of the potential with $\tau_i$ is
exponentially suppressed. We can find the global minimum by
extremising (\ref{eqSPT}) with respect to $\tau_i$. Doing this at
fixed $\mc{V}$, we obtain \be (a_i A_i) e^{-a_i \tau_i} = \frac{3
\alpha \lambda_i W_0}{2 \mc{V}} \frac{(1 - a_i \tau_i)}{(\half - 2
a_i \tau_i)} \sqrt{\tau_i}. \ee If we approximate $a_i \tau_i >>
1$ (which is valid at large volume as $a_i \tau_i \sim \ln
(\mc{V}))$, then substituting this into the potential
(\ref{eqSPT}) contributes
$\frac{-3 \lambda_i W_0^2}{2 \mc{V}^3} \tau_{i,min}^{3/2} \alpha,$
 which can be reexpressed as
 $ \label{eqMCP} \frac{-3
\lambda_i W_0^2 \alpha}{2 \mc{V}^3 a_i^{3/2}} (\ln \mc{V} -
c_i)^{3/2},$
where $c_i = \ln (\frac{3 \alpha \lambda_i W_0}{2
a_i A_i})$. At large values of $\ln \mc{V}$, the resulting
potential for the volume once all $\tau_i$ fields are minimised is
\be V = \frac{-3 W_0^2}{2 \mc{V}^3} \left( \sum_{i=2}^n \left[
\frac{\lambda_i
  \alpha}{a_i^{3/2}} \right] (\ln \mc{V})^{3/2} -
\frac{\xi}{2} \right). \ee This is the reason why the volume may
be exponentially large. It is necessary to add an uplift term to
ensure that the minimum is essentially Minkowksi. For concreteness
we use IASD fluxes\footnote{These are pure supergravity and so it is 
  manifest that the uplift only depends on the volume with
  unwanted dependence on $\tau_i$.} and write the volume potential as
\be
\label{eqPUP}
V = \frac{-3 W_0^2}{2 \mc{V}^3} \left( \sum_{i=2}^n \left[ \frac{\lambda_i
  \alpha}{a_i^{3/2}} \right] (\ln \mc{V})^{3/2} -
\frac{\xi}{2} \right) + \frac{\gamma W_0^2}{\mc{V}^2}, \ee where
$\gamma \sim \mc{O}(\frac{1}{\mc{V}})$ parametrises the magnitude
of the uplift. By tuning $\gamma$, the potential (\ref{eqPUP})
(and by extension its full form (\ref{eqSPT})) has a Minkowski or
small de Sitter minimum.

To obtain inflation we consider the potential away from the minimum.
We take a `small' modulus, say $\tau_n$, as the
inflaton and displace it far from its minimum. At constant volume the potential is
exponentially flat along this direction, and the modulus
rolls back in an inflationary fashion.
There is no problem in terms of initial conditions.
While we do not know how the moduli evolution
starts, we do know how it must end, namely with all moduli at their
minima. Given this - we have nothing new to say on the
overshoot problem \cite{hepth9212049} - inflation occurs as the last K\"ahler modulus
rolls down to its minimum.

It is necessary that all other moduli, and in particular the volume, are
stable during inflation.
Displacing $\tau_n$ from its minimum nullifies the contribution
made by the stabilised $\tau_n$ to the volume potential.
The effective volume potential during inflation is then
\be
\label{eqMPI}
V = \frac{-3 W_0^2}{2 \mc{V}^3} \left( \sum_{i=2}^{n-1} \left[ \frac{\lambda_i
  \alpha}{a_i^{3/2}} \right] (\ln \mc{V})^{3/2} -
\frac{\xi}{2} \right) + \frac{\gamma W_0^2}{\mc{V}^2}, \ee
Provided that the ratio
 \be  \label{eqRAT}
\rho\ \equiv \ \frac{\lambda_n}{a_n^{3/2}} \, : \, \sum_{i=2}^{n}
\frac{\lambda_i}{a_i^{3/2}} \ee is sufficiently
small\footnote{This can be quantified in explicit models. For
large volumes the condition on the ratio $\rho$ is that $9.5
(\ln \mc{V} ) \rho  < 1$. 
As long as we restrict to reasonable values for the $a_i$,
this bounds the volume
  at the minimum.
To obtain inflation with correct density perturbations, the
  appropriate volumes are
$\mc{O}(10^5 - 10^7)$, which
can be satisfied using sensible values for
  $\lambda_i$ and $a_i$.}, there is little
difference between
  (\ref{eqPUP}) and (\ref{eqMPI}) and the volume modulus will be stable
during inflation.  
As we obviously require $\rho < 1$, it follows that at least three K\"ahler
moduli are necessary.
While (\ref{eqRAT}) can always be satisfied by an appropriate choice of $a_i$, this
  becomes easier and easier with more K\"ahler moduli.

\begin{figure}[ht]
\linespread{0.2}
\begin{center}
\makebox[16cm]{ \epsfxsize=16cm \epsfysize=10cm \epsfbox{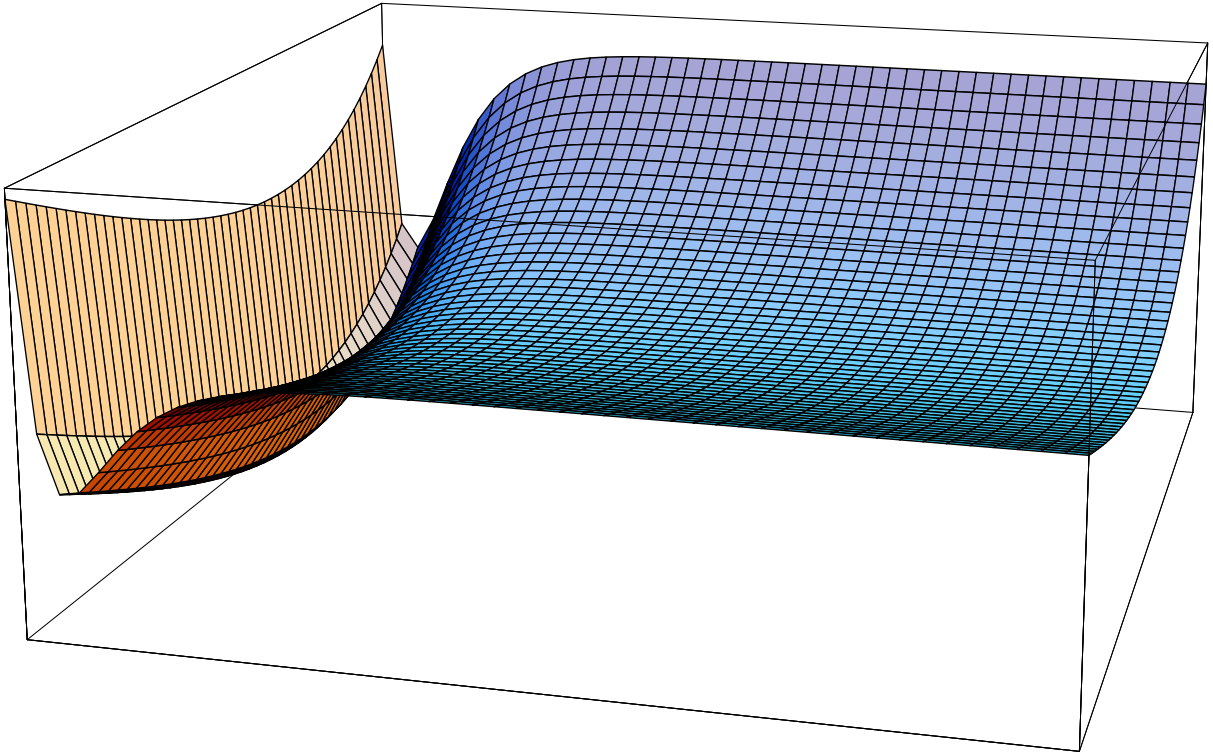}}
\end{center}
\caption{ Inflationary potential: the inflaton lies along the
  x-direction and the volume along the y-direction.}
\label{picIP}
\end{figure}

We illustrate the form of the resulting inflationary potential in figure
\ref{picIP}, showing the inflaton and volume directions.

\section{Inflationary Potential and Parameters}

Let us now quantify the resulting potential and compute the
inflationary parameters. The inflationary potential is
read off from (\ref{eqSPT}) to be \be V_{inf} = V_0 - \frac{4
\tau_n W_0 a_n A_n e^{-a_n \tau_n}}{\mc{V}^2}, \ee as the double
exponential in (\ref{eqSPT}) is irrelevant during inflation.
During inflation $V_0$ is constant and can be parametrised as \be
V_0 = \frac{\beta W_0^2}{\mc{V}^3}. \ee ($\frac{1}{\mc{V}^3}$ is the
scale of the potential during inflation). However, $\tau_n$ is not
canonically normalised, as to leading order in volume \be K_{n
\bar{n}} = \frac{3 \lambda}{8 \sqrt{\tau_n} \mc{V}}. \ee The
canonically normalised field is \be \tau_n^{c} = \sqrt{\frac{4
\lambda}{3 \mc{V}}} \tau_n^{\frac{3}{4}}. \ee
In terms of
$\tau_n^{c}$, the inflationary potential is
\be
V = V_0 - \frac{4
W_0 a_n A_n}{\mc{V}^2}
\left(\frac{3 \mc{V}}{4 \lambda} \right)^{2/3}  (\tau_n^{c})^{4/3}
\exp \left[-a_n \left(\frac{3 \mc{V}}{4
    \lambda}\right)^{2/3} (\tau_n^{c})^{4/3}\right].
\ee
This is similar, but not identical, to the textbook potential
$V = V_0 (1 - e^{-\tau})$. Although $\tau_n^c$ is canonically
normalised, it has no natural geometric interpretation and for clarity
we shall express the inflationary
parameters in terms of $\tau_n$, the cycle volume.

The slow-roll parameters are defined by
\bea
\epsilon & = & \frac{M_P^2}{2} \left( \frac{V'}{V} \right)^2, \\
\eta & = & M_P^2 \frac{V''}{V}, \\
\xi & = & M_P^4 \frac{V' V^{'''}}{V^2},
\eea
with the derivatives being with respect to $\tau_n^c$.
These can be evaluated to give
\bea
\epsilon & = & \frac{32 \mc{V}^3}{3 \beta^2 W_0^2} a_n^2 A_n^2
\sqrt{\tau_n} (1 - a_n \tau_n)^2 e^{-2 a_n \tau_n}, \nonumber \\
\eta & = & - \frac{4 a_n A_n \mc{V}^2}{3 \lambda \sqrt{\tau_n} \beta W_0}
\left[ (1 - 9 a_n \tau_n + 4 (a_n \tau_n)^2) e^{-a_n \tau_n} \right],
\\
\xi & = & \frac{- 32 (a_n A_n)^2 \mc{V}^4}{9 \beta^2 \lambda^2 W_0^2 \tau_n}
 (1 - a_n \tau_n) (1 + 11 a_n \tau_n - 30 (a_n \tau_n)^2 + 8
(a_n \tau_n)^2) e^{-2 a_n \tau_n}.\nonumber
 \eea
 Then $\xi <<
\epsilon, \eta << 1$ provided that $e^{-a_n
  \tau_n} << \frac{1}{\mc{V}^2}$.

Within the slow-roll approximation, the spectral index and its
running are given by
\bea
\label{eqSIR}
n - 1 & = & 2 \eta - 6 \epsilon + \mc{O}(\xi), \\
\frac{d n}{d \ln k} & = & 16 \epsilon \eta - 24 \epsilon^2 - 2 \xi.
\eea
The number of efoldings is given by
\be
N_e = \int_{\phi_{end}}^{\phi} \frac{V}{V'} d \phi,
\ee
which may be expressed as
\be
N_e = \frac{-3 \beta W_0 \lambda_n}{16 \mc{V}^2 a_n A_n}
\int_{\tau_n^{end}}^{\tau_n} \frac{e^{a_n \tau_n}}{\sqrt{\tau_n} (1 -
  a_n \tau_n)} d \tau_n.
\ee Matching the COBE normalisation for the density fluctuations
$\delta_H = 1.92 \times 10^{-5} $ requires \be \label{eqCNF}
\frac{V^{3/2}}{M_P^3 V'} = 5.2 \ti 10^{-4}, \ee where the LHS is
evaluated at horizon exit, $N_e = 50 - 60$ efoldings before the
end of inflation. This condition can be expressed as \be
\label{eqDP} \left( \frac{g_s^4}{8\pi} \right) \frac{3 \lambda \beta^3 W_0^2}{64 \sqrt{\tau_n} (1 -
a_n \tau_n)^2} \left( \frac{W_0}{a_n A_n} \right)^2 \frac{e^{2 a_n
\tau_n}}{
    \mc{V}^6} = 2.7 \ti 10^{-7}.
\ee
We have here included a factor of $\frac{g_s^4}{8 \pi}$ that should
properly be included as an overall normalisation in $V$ - see \cite{hepth0505076}.
The condition (\ref{eqCNF}) determines the normalisation of the potential and
in practice we use it as a constraint on the stabilised volume.

Finally, the tensor-to-scalar ratio is
\be
\label{eqTSR}
r \sim 12.4 \epsilon.
\ee

\subsection{Footprint of the Model}

We now want to determine the inflationary predictions for the above model.
In the above model
there are various undetermined
parameters arising from the detailed microphysics, such as the
threshold correction $A$ or tree-level superpotential $W_0$. In
principle, these are determined by the specific Calabi-Yau with its brane
and flux configurations, but they can be prohibitively difficult to
calculate in realistic examples.
However, it turns out that the most important results are independent
of these parameters.
In particular, solving equations (\ref{eqSIR}) to (\ref{eqTSR})
numerically,
we find the robust results
\bea
\eta & \approx & -\frac{1}{N_e}, \\
\epsilon & < & 10^{-12}, \\
\xi & \approx & -\frac{2}{N_e^2}.
\eea
These results are not so surprising given the similarity of the
potential to the textbook form $V_0 ( 1 - e^{-\tau})$.
Taking a range of $N_e = 50 \to 60$, we obtain in the slow-roll approximation
\bea
0.960 & < n < & 0.967, \\
-0.0006 & < \frac{d n}{d \ln k} < & -0.0008, \\
0 & < \vert r \vert < & 10^{-10},
\eea
where the uncertainties above arise principally from the number of e-foldings. If we go beyond the
slow-roll approximation, the expression for $n$ will receive
$\mc{O}(\xi)$ corrections - these are minimal and can be neglected.

To evaluate the inflationary energy scale, it is convenient to
reformulate the COBE normalisation of density
perturbations $\delta_H
= 1.92 \times 10^{-5} $ as \be \frac{V^{1/4}}{\epsilon^{1/4}} =
6.6 \ti 10^{16} \textrm{GeV}. \ee
Unlike the predictions for the spectral index, the required internal
volume is parameter-dependent.
For typical values of the microscopic parameters 
this is found numerically to take a range of values \be 10^5 l_s^6
\le \mc{V} \le 10^7 l_s^6, \ee where $l_s = (2 \pi)
\sqrt{\alpha'}$. As the moduli stabilisation mechanism of
\cite{hepth0502058, hepth0505076} naturally generates
exponentially large volumes, there is no difficulty in achieving
these values. The range of $\epsilon$ at horizon exit
is $10^{-13} \ge \epsilon \ge 10^{-15}$, and thus the
inflationary energy scale is rather low, \be V_{inf} \sim 10^{13}
\textrm{GeV}. \ee This implies in particular that tensor
perturbations would be unobservable in this model.

There is no practical upper limit on the number of efoldings attainable. This is
large-field inflation and the potential is exponentially flat as the
inflaton 4-cycle increases in volume. A very large number of efoldings is
achieved by a very small variation in the inflaton and barring
cancellations we would expect $N_{e,total} >> 60$ in these models.

In these compactifications, the lightest non-axionic modulus has a
mass \cite{hepth0505076}
\be
M \sim \frac{M_P}{\mc{V}^{3/2}}.
\ee
Thus even at the larger end of volumes $M >> \mc{O}(10) \textrm{TeV}$ and
there is no cosmological moduli problem. As indicated earlier, there
is also not a problem with initial conditions for inflation. Given that the moduli
attain their minimum, the inflaton is simply the last K\"ahler modulus
to roll down to the minimum. We do not need to worry about
interference from the evolution of the other moduli. Once they roll
down to the minimum they become heavy and will rapidly decouple from
inflationary dynamics.

We have nothing new to say on the cosmological overshoot
problem. It is difficult to see how progress may be achieved here
without an adequate formulation of initial conditions for the
universe. This problem is amplified by the fact that typical
Calabi-Yaus have hundreds of both complex structure and K\"ahler
moduli; it is very difficult to give a well-motivated choice for the initial values
and evolution of so many moduli. (For the possibility that damping can
remove the overshoot problem see \cite{KaloperOlive, hepth0409226, brustein}).

\subsection{Additional Corrections and Extensions}

The inflationary mechanism presented here relies on the exponential
flatness of the $\tau_n$ direction at constant volume. This is
unbroken by the tree-level K\"ahler potential, the $(\alpha')^3$
correction and the uplift term.
Let us briefly discuss effects
that might spoil this.

Let us first focus on superpotential effects. The nonrenormalisation
theorems guarantee that the K\"ahler moduli cannot appear
perturbatively in $W$.
However, the flatness could be spoiled if the gauge kinetic functions
$A_i$ depended polynomially on the K\"ahler moduli.
A term $A(T_j) e^{-T_i}$ in the superpotential would lead to an
effective polynomial term for $T_j$ once $T_i$ was stabilised.
However, the $A_i$
must be holomorphic in $T_i$ and respect the axion shift symmetries,
and so this polynomial dependence on $T_i$ cannot occur. Indeed, in models
for which the threshold corrections have been computed
explicitly, there is no dependence of the gauge kinetic functions on the K\"ahler moduli
\cite{berghaackkors}. Combined with non-renormalisation results, this
means that the exponential flatness cannot be lifted by
superpotential effects.

The other possibility is that the exponential flatness may be lifted
by corrections to
the K\"ahler potential that depend 
on $\tau_n$.
Considering first bulk terms, both
the tree-level K\"ahler potential and the $\mc{O}(\alpha'^3)$
correction computed in \cite{hepth0204254} have the property that
their contribution to the scalar potential is only a function of the
volume and has no explicit dependence on the moduli. These then do not
affect the constant volume flatness of the $\tau_n$ direction. It would be
interesting, but difficult, to determine whether this feature extends beyond
the terms so far computed.

There are also open string K\"ahler corrections such as those recently computed in
\cite{berghaackkors}. Of necessity, this computation is restricted to
certain toroidal orientifolds with D3 and D7 branes. The 1-loop 
corrections determined there are subdominant in the scalar
potential to the
$\mc{O}(\alpha'^3)$ corrections,
although they give a larger contribution
to the K\"ahler potential. This counterintuitive result is due to the
fact that 
a K\"ahler correction
$$
K + \delta K = - 2 \ln (\mc{V}) + \frac{\epsilon}{\mc{V}^{2/3}}
$$
only gives in the scalar potential
$$
\frac{\delta V}{V} = \frac{\mc{O}(\epsilon)}{\mc{V}^{4/3}}.
$$
Thus $\mc{O}(\mc{V}^{-2/3})$ corrections to the K\"ahler potential are
in fact subdominant in the scalar potential to $\mc{O}(\mc{V}^{-1})$ corrections.
This result makes the volume stabilisation
mechanism of \cite{hepth0502058, hepth0505076} more robust.
For the models for which the string loop computation can be performed,
there is no analogue
of the $\tau_n$ blow-up field and so it is
unclear whether and in what fashion these might
appear in the 1-loop correction to the K\"ahler potential.
There are however physical constraints: $e^K$ appears in the scalar
potential and so must behave sensibly in the limits of both small and
large $\tau_n$.

There are also field theory loop corrections determined in
\cite{hepth0507131}.
These are again subdominant in the scalar potential to the
$\mc{O}(\alpha'^3)$ corrections used above for volume stabilisation. 
This computation again does not have an
analogue of the blow-up modes we have used for the inflaton.

The upshot is that the exponential flatness of the
$\tau_n$ direction is not broken by any of the known corrections. 
In general, any correction that can be expressed in terms of the
overall volume will not alter the exponential flatness of the $\tau_n$ direction.
If
corrections existed which did break this exponential flatness, it
would be necessary to examine their form and magnitude - it is not after all
necessary that the exponential flatness survive for all values of
$\tau_n$, but merely for those relevant during the last sixty e-folds.

Finally, we have used an oversimplified form for the Calabi-Yau,
picturing it as simply a combination of a volume cycle and blow-up modes.
This is not necessary for the inflationary mechanism described
here. 
Whilst in (\ref{vol}) we assumed $h^{1,1} - 1$ moduli
to be blow-ups whose only nonvanishing triple intersection was with
themselves, 
a single such modulus would be perfectly adequate as an inflaton.
Indeed, even this is not necessary - 
the minimal requirement is simply a flat
direction, which originates from the no-scale behaviour and is broken by
nonperturbative effects. The condition necessary to ensure the volume is stable
during inflaton will then be a generalisation of (\ref{eqRAT}).

\section{Discussion}

We have presented a general but simple scenario of inflation in
string theory that does not require fine tuning of parameters,
applies to a very large class of compactifications and is
predictive at the level that can be ruled out within a few years.
This scenario realises large field inflation in a natural way. The
main properties of these models are the existence of flat
directions broken by non-perturbative effects. The flat directions
have their origins in the no-scale property of the K\"ahler
potential and are generic for IIB K\"ahler moduli, as is the
appearance of instanton-generated nonperturbative superpotentials.
The scenario is embedded in the exponentially large volume
compactifications of \cite{hepth0502058, hepth0505076} and
requires $h^{2,1} > h^{1,1}$ and $h^{1,1} > 2$. This last
requirement is necessary to ensure that the volume is stabilised
during inflation.

Notice that the volumes required to obtain inflation, while large, are not
extremely large as the string scale is only a few
orders of magnitude below the Planck scale. The necessary volumes
of $\mc{O}(10^5-10^7)$ in string units can be obtained by natural choices of
the exponential parameters $a_i$ ($a\sim
\frac{2 \pi}{3}$ in
the simplest cases) \cite{hepth0505076}.

Although there are many moduli, the inflationary period reduces to a
single-field case. This is because the inflaton is simply the last modulus to
roll down to its minimum, and once other moduli attain their minimum
they rapidly become heavy and decouple from inflationary dynamics.
In principle there
are at least two other fields that may have a nontrivial role during
the cosmological evolution. One is the axion partner of the
inflaton field. We have chosen this to sit at the minimum of its
oscillatory potential, at least for the last sixty efolds. This is
not a strong assumption - because the inflaton
direction is so flat, there is a lot of time for the axion to relax 
from a possibly non-zero inital value to its minimum
before the last sixty efolds start.
(Remember that depending on initial conditions
the total number of efolds may be many orders of magnitude
larger than 60.) It would nonetheless be interesting to study a
multiple field inflation configuration in which both fields
contribute to the density perturbations \cite{bartjan}. 

There is also a
second direction which is extremely flat, corresponding
to the axionic partner of the overall volume modulus. This field
is so light, with $m << 10^{-300}$ GeV
\cite{hepth0505076} both during and after inflation, that it will not play a role in the
cosmological evolution.

We have considered inflation as occurring at the top of a waterfall,
and inflation ending as the moduli roll down to the waterfall. It may
also be interesting to consider the case where there are multiple
waterfalls. By appropriately tuning the uplift we could arrange that
the current vacuum energy corresponds to the top of a waterfall rather
than the bottom. As the field would be slowly rolling this would then
correspond to quintessence.

Another open question concerns reheating. Note
that unlike brane inflation, in which reheating is driven
by tachyon condensation \cite{hepth0105204} requiring non-trivial
string theory dynamics to be understood \cite{reheating}, in our
case, as in racetrack inflation, reheating is a pure
field theory problem that only requires the study of the
matter/inflaton couplings. In this respect, if the
standard model lies on D7 branes wrapping the four-cycle
whose size is determined by the inflaton field, the inflaton
can decay directly to the  gauge fields of the standard model
through the coupling $\tau_n F^{\mu\nu} F_{\mu\nu}$. This can give
rise to efficient reheating as discussed in e.g.
\cite{lindebook}. If the standard model lies elsewhere,
the inflaton will couple to standard model fields through
higher dimension operators and a more detailed analysis is required.
(For a recent analysis of reheating in brane-antibrane inflation see \cite{hepth0508229}).

It is worth comparing aspects of this mechanism with other
inflationary models obtained from string theory. Besides the issue
of fine tuning, it differs from racetrack inflation and tachyon
driven inflation in that it corresponds to large field rather than
hill-top inflation. 
In principle our scenario is closer to brane separation
inflation, but we do not need  a second field to end inflation and
in particular do not predict the existence of remnant cosmic
strings from the reheating era. Numerically, our predictions are close to racetrack
inflation although with the spectral index within a more
comfortable range. The exponentially flat direction resembles the
mechanism of \cite{hepth0111025}, with the advantage
that moduli stabilisation is now derived and not assumed.

Let us finally discuss the generality of our scenario. 
The main technical assumption we
have used is the direct expression for the volume in terms of the
K\"ahler moduli (\ref{vol}). This was overkill - the only part of the 
assumption we actually used
was that the inflaton modulus appears alone in the volume as
${\mc V}= \ldots -(T_n+\bar{T}_n)^{\frac{3}{2}}$. 
As indicated above, we can relax even this: the absolute minimal
requirement is simply the existence of a flat direction broken by
nonperturbative effects.
There may be several possible inflationary directions - in the above
model, $\tau_2, \ldots, \tau_n$ are all good candidates - with
the particular one chosen determined by which K\"ahler modulus is 
last to attain its minimum.
In each case we expect similar
physics to emerge with a robust prediction on the spectral index of density
perturbations. It is very exciting that such a simple string
scenario has the basic properties needed for a 
realisation of cosmological inflation with predictions that can be
confirmed or ruled out in the  near future.

\section*{Acknowledgements}
We acknowledge useful conversations with P. Berglund, C. Burgess,
S. Kachru and B. van Tent. The research of FQ is partially funded
by PPARC and a Royal Society Wolfson merit award. JC is grateful to
EPSRC for a research studentship.


\begin{thebibliography}{99}
\bibitem{hepth0105097}
S.~B.~Giddings, S.~Kachru and J.~Polchinski,
``Hierarchies from fluxes in string compactifications,''
Phys.\ Rev.\ D {\bf 66}, 106006 (2002)
[arXiv:hep-th/0105097].

\bibitem{hepth0301240}
S.~Kachru, R.~Kallosh, A.~Linde and S.~P.~Trivedi,
``De Sitter vacua in string theory,''
Phys.\ Rev.\ D {\bf 68}, 046005 (2003)
[arXiv:hep-th/0301240].

\bibitem{WMAP}
C.~L.~Bennett {\it et al.},
``First Year Wilkinson Microwave Anisotropy Probe (WMAP) Observations:
Preliminary Maps and Basic Results,''
Astrophys.\ J.\ Suppl.\  {\bf 148} (2003) 1
[arXiv:astro-ph/0302207].

\bibitem{bg}P.~Binetruy and M.~K.~Gaillard,
``Candidates For The Inflaton Field In Superstring Models,''
Phys.\ Rev.\ D {\bf 34} (1986) 3069;
T.~Banks, M.~Berkooz, S.~H.~Shenker, G.~W.~Moore and
P.~J.~Steinhardt, ``Modular cosmology,'' Phys.\ Rev.\ D {\bf 52}
(1995) 3548 [arXiv:hep-th/9503114].

\bibitem{hepph9812483}
G.~R.~Dvali and S.~H.~H.~Tye,
``Brane inflation,''
Phys.\ Lett.\ B {\bf 450}, 72 (1999)
[arXiv:hep-ph/9812483].

\bibitem{hepth0105204}
C.~P.~Burgess, M.~Majumdar, D.~Nolte, F.~Quevedo, G.~Rajesh and
R.~J.~Zhang, ``The inflationary brane-antibrane universe,'' JHEP
{\bf 0107}, 047 (2001) [arXiv:hep-th/0105204].

\bibitem{dss}
G.~R.~Dvali, Q.~Shafi and S.~Solganik,
  ``D-brane inflation,''
  arXiv:hep-th/0105203.

\bibitem{hepth0111025}
C.~P.~Burgess, P.~Martineau, F.~Quevedo, G.~Rajesh and R.~J.~Zhang,
``Brane antibrane inflation in orbifold and orientifold models,''
JHEP {\bf 0203}, 052 (2002)
[arXiv:hep-th/0111025].

\bibitem{hepth0406230}
J.~J.~Blanco-Pillado {\it et al.},
``Racetrack inflation,''
JHEP {\bf 0411}, 063 (2004)
[arXiv:hep-th/0406230].

\bibitem{hepth0308055}
S.~Kachru, R.~Kallosh, A.~Linde, J.~Maldacena, L.~McAllister and
S.~P.~Trivedi, ``Towards inflation in string theory,'' JCAP {\bf
0310} (2003) 013 [arXiv:hep-th/0308055].

\bibitem{renata}
J.~P.~Hsu, R.~Kallosh and S.~Prokushkin, ``On brane inflation with
volume stabilization,'' JCAP {\bf 0312} (2003) 009
[arXiv:hep-th/0311077];
F.~Koyama, Y.~Tachikawa and T.~Watari, ``Supergravity analysis of
hybrid inflation model from D3-D7 system'',
[arXiv:hep-th/0311191];
H.~Firouzjahi and S.~H.~H.~Tye, ``Closer towards inflation in
string theory,'' Phys.\ Lett.\ B {\bf 584} (2004) 147
[arXiv:hep-th/0312020].
J.~P.~Hsu and R.~Kallosh, ``Volume stabilization and the origin of
the inflaton shift symmetry in string theory,'' JHEP {\bf 0404}
(2004) 042 [arXiv:hep-th/0402047];

\bibitem{various}
C.~P.~Burgess, J.~M.~Cline, H.~Stoica and F.~Quevedo, ``Inflation
in realistic D-brane models,'' [arXiv:hep-th/0403119];
J.~M.~Cline and H.~Stoica,
  ``Multibrane inflation and dynamical flattening of the inflaton potential,''
  arXiv:hep-th/0508029.
O.~DeWolfe, S.~Kachru and H.~Verlinde, ``The giant inflaton,''
JHEP {\bf 0405} (2004) 017 [arXiv:hep-th/0403123].
N.~Iizuka and S.~P.~Trivedi, ``An inflationary model in string
theory,'' arXiv:hep-th/0403203.
K.~Becker, M.~Becker and A.~Krause,
  ``M-theory inflation from multi M5-brane dynamics,''
  Nucl.\ Phys.\ B {\bf 715} (2005) 349
  [arXiv:hep-th/0501130].
D.~Cremades, F.~Quevedo and A.~Sinha,
``Warped tachyonic inflation in type IIB flux compactifications and the
open-string completeness conjecture,''
arXiv:hep-th/0505252.
S.~Dimopoulos, S.~Kachru, J.~McGreevy and J.~G.~Wacker,
``N-flation,''
arXiv:hep-th/0507205.
E.~Silverstein and D.~Tong,
``Scalar speed limits and cosmology: Acceleration from D-cceleration,''
Phys.\ Rev.\ D {\bf 70} (2004) 103505
[arXiv:hep-th/0310221];
M.~Alishahiha, E.~Silverstein and D.~Tong,
``DBI in the sky,''
Phys.\ Rev.\ D {\bf 70} (2004) 123505
[arXiv:hep-th/0404084];
X.~g.~Chen,
``Inflation from warped space,''
arXiv:hep-th/0501184;
H.~Singh,
``(A)symmetric tachyon rolling in de Sitter spacetime: A universe devoid of
Planck density,''
arXiv:hep-th/0508101.


\bibitem{hepth0502058}
V.~Balasubramanian, P.~Berglund, J.~P.~Conlon and F.~Quevedo,
``Systematics of moduli stabilisation in Calabi-Yau flux compactifications,''
JHEP {\bf 0503} (2005) 007
[arXiv:hep-th/0502058].


\bibitem{hepth0505076}
J.~P.~Conlon, F.~Quevedo and K.~Suruliz,
``Large-volume flux compactifications: Moduli spectrum and D3/D7 soft
supersymmetry breaking,''
arXiv:hep-th/0505076.

\bibitem{LiddleLyth}
A.~R.~Liddle and D.~H.~Lyth,
``Cosmological inflation and large-scale structure,''
CUP (2000).

\bibitem{hepth0309187}
C.~P.~Burgess, R.~Kallosh and F.~Quevedo,
``de Sitter string vacua from supersymmetric D-terms,''
JHEP {\bf 0310}, 056 (2003)
[arXiv:hep-th/0309187].

\bibitem{hepth0402135}
A.~Saltman and E.~Silverstein,
JHEP {\bf 0411}, 066 (2004)
[arXiv:hep-th/0402135].

\bibitem{hepth0204254}
K.~Becker, M.~Becker, M.~Haack and J.~Louis,
``Supersymmetry breaking and alpha'-corrections to flux induced  potentials,''
JHEP {\bf 0206}, 060 (2002)
[arXiv:hep-th/0204254].

\bibitem{berghaackkors}
M.~Berg, M.~Haack and B.~Kors,
``Loop corrections to volume moduli and inflation in string theory,''
Phys.\ Rev.\ D {\bf 71}, 026005 (2005)
[arXiv:hep-th/0404087].
M.~Berg, M.~Haack and B.~Kors,
``String Loop Corrections to Kahler Potentials in Orientifolds,''
arXiv:hep-th/0508043.
M.~Berg, M.~Haack and B.~Kors,
``On Volume Stabilization by Quantum Corrections,''
arXiv:hep-th/0508171.

\bibitem{hepth0507131}
G.~von Gersdorff and A.~Hebecker,
``Kaehler corrections for the volume modulus of flux compactifications,''
arXiv:hep-th/0507131.

\bibitem{per}
V.~Balasubramanian and P.~Berglund,
``Stringy corrections to Kaehler potentials, SUSY breaking, and the
cosmological constant problem,''
JHEP {\bf 0411}, 085 (2004)
[arXiv:hep-th/0408054].
P.~Berglund and P.~Mayr,
``Non-perturbative superpotentials in F-theory and string duality,''
arXiv:hep-th/0504058.

\bibitem{bw}
K.~Bobkov,
``Volume stabilization via alpha' corrections in type IIB theory with fluxes,''
JHEP {\bf 0505}, 010 (2005)
[arXiv:hep-th/0412239].
A.~Westphal,
``Eternal inflation with alpha' corrections,''
arXiv:hep-th/0507079.


\bibitem{CandelasDeLaOssa}
P.~Candelas and X.~de la Ossa,
``Moduli Space Of Calabi-Yau Manifolds,''
Nucl.\ Phys.\ B {\bf 355}, 455 (1991).

\bibitem{hepth0404257}
F.~Denef, M.~R.~Douglas and B.~Florea,
``Building a better racetrack,''
JHEP {\bf 0406}, 034 (2004)
[arXiv:hep-th/0404257].

\bibitem{hepth9212049}
R.~Brustein and P.~J.~Steinhardt,
``Challenges for superstring cosmology,''
Phys.\ Lett.\ B {\bf 302}, 196 (1993)
[arXiv:hep-th/9212049].

\bibitem{KaloperOlive}
N.~Kaloper and K.~A.~Olive,
``Dilatons in string cosmology,''
Astropart.\ Phys.\  {\bf 1} (1993) 185.

\bibitem{hepth0409226}
N.~Kaloper, J.~Rahmfeld and L.~Sorbo,
``Moduli entrapment with primordial black holes,''
Phys.\ Lett.\ B {\bf 606}, 234 (2005)
[arXiv:hep-th/0409226].

\bibitem{brustein}
R.~Brustein, S.~P.~de Alwis and P.~Martens,
  ``Cosmological stabilization of moduli with steep potentials,''
  Phys.\ Rev.\ D {\bf 70} (2004) 126012
  [arXiv:hep-th/0408160].

\bibitem{bartjan}
B.~van Tent,
  ``Multiple-field inflation and the CMB,''
  Class.\ Quant.\ Grav.\  {\bf 21} (2004) 349
  [arXiv:astro-ph/0307048].



\bibitem{reheating}
G.~Shiu, S.~H.~H.~Tye and I.~Wasserman,
  ``Rolling tachyon in brane world cosmology from superstring field theory,''
  Phys.\ Rev.\ D {\bf 67} (2003) 083517
  [arXiv:hep-th/0207119];
J.~M.~Cline, H.~Firouzjahi and P.~Martineau,
  ``Reheating from tachyon condensation,''
  JHEP {\bf 0211} (2002) 041
  [arXiv:hep-th/0207156];
N.~Barnaby, C.~P.~Burgess and J.~M.~Cline,
  ``Warped reheating in brane-antibrane inflation,''
  JCAP {\bf 0504} (2005) 007
  [arXiv:hep-th/0412040];
L.~Kofman and P.~Yi,
  ``Reheating the universe after string theory inflation,''
  arXiv:hep-th/0507257.
A.~R.~Frey, A.~Mazumdar and R.~Myers,
  ``Stringy Effects During Inflation and Reheating,''
  arXiv:hep-th/0508139.


\bibitem{lindebook}
A.~D.~Linde,
  ``Particle Physics and Inflationary Cosmology,''
Harwood (1990) 362 p. (Contemporary concepts in physics, 5)
  arXiv:hep-th/0503203.

\bibitem{hepth0508229}
D.~Chialva, G.~Shiu and B.~Underwood,
``Warped reheating in multi-throat brane inflation,''
arXiv:hep-th/0508229.

\end{thebibliography}
\end{document}